\documentclass[runningheads]{llncs}
\usepackage{enumitem}
\usepackage{amsmath} 
\usepackage{tabularx} % 在文档导言区添加这个宏包
\usepackage{arydshln} % 添加这一行来使用虚线命令
\usepackage{marvosym}
\usepackage{hyperref}
\usepackage[T1]{fontenc}
\PassOptionsToPackage{colorlinks=true,linkcolor=blue,urlcolor=blue,citecolor=blue}{hyperref}
\usepackage{hyperref}
\usepackage[colorlinks=true, linkcolor=blue, urlcolor=blue, citecolor=blue]{hyperref}
\usepackage{float}
\usepackage{amssymb}  % 导入amssymb包

\usepackage{graphicx}
\begin{document}

\title{PCQ: Emotion Recognition in Speech via Progressive Channel Querying}
\author{Xincheng Wang$^1$ \and Liejun Wang$^1$\textsuperscript{(\Letter)} \and Yinfeng Yu$^1$\textsuperscript{(\Letter)} \and Xinxin Jiao$^1$}

\authorrunning{Xincheng Wang et al.}
\institute{$^1$School of Computer Science and Technology, Xinjiang University, China
\email{wljxju,yuyinfeng}@xju.edu.cn}
%\email{\{wljxju,yuyinfeng\}@uni-heidelberg.de}}
\maketitle          % typeset the header of the contribution%
\begin{abstract}
In human-computer interaction (HCI), Speech Emotion Recognition (SER) is a key technology for understanding human intentions and emotions. Traditional SER methods struggle to effectively capture the long-term temporal correla-tions and dynamic variations in complex emotional expressions. To overcome these limitations, we introduce the PCQ method, a pioneering approach for SER via \textbf{P}rogressive \textbf{C}hannel \textbf{Q}uerying. This method can drill down layer by layer in the channel dimension through the channel query technique to achieve dynamic modeling of long-term contextual information of emotions. This mul-ti-level analysis gives the PCQ method an edge in capturing the nuances of hu-man emotions. Experimental results show that our model improves the weighted average (WA) accuracy by 3.98\% and 3.45\% and the unweighted av-erage (UA) accuracy by 5.67\% and 5.83\% on the IEMOCAP and EMODB emotion recognition datasets, respectively, significantly exceeding the baseline levels. 

\end{abstract}
\section{Introduction}
Emotion recognition technology identifies and understands an individual's emotional state by analyzing multiple modalities\cite{c1}, including speech, facial expres-sions, body movements, and language. This technology is essential in some fields, such as Face Forgeries\cite{c2} and health monitoring \cite{c3}. Especially in specific scenarios, like call centres and telemedicine, speech becomes the preferred or the only feasible modal-ity for emotion recognition due to other modalities' unavailability or practical limita-tions, such as text and images. This modal dependence highlights speech emotion recognition's unique value and challenges in specific applications. Given this, this paper will investigate unimodal speech emotion recognition.

Much attention has been paid to the temporal nature of speech signals \cite{c4}, a crucial property for understanding and processing speech data. Our research focuses on discrete speech emotion recognition (DSER) \cite{c5}, i.e., in this framework, we assume that each sentence expresses a single emotion. However, the expression of human emo-tions is a dynamic process: emotions are not fully revealed instantly but gradually revealed over time. This dynamism means that although our task is to identify the dominant emotion in each sentence, we still need to pay attention to temporal varia-tions in the speech signal to capture subtle clues about how the emotion develops over time. The temporal character of emotional expression exacerbates the difficulty of accurately dynamic modelling emotional information in speech. To address this chal-lenge, current research approaches face several problems. Firstly, although the Transformer \cite{c6} based approach enables global modelling, the approach may not be appli-cable when dealing with relatively small unimodal speech emotion recognition data. Therefore, current research in SER focuses mainly on capturing contextual emotional information using self-attention or cross-attention mechanisms. Hu et al. in \cite{c7} applied different cross-attention modules to a joint speech emotion recognition network and achieved good results in a speaker-independent setting. In \cite{c8}, Jiao et al. proposed a Hierarchical Cooperative Attention method, which combines features extracted by HuBERT with spectrogram features to enhance the accuracy of speech emotion recognition. In addition, Xu et al. proposed a head fusion strategy in \cite{c9}, which solves the limitation that the multi-head attention mechanism can only focus on a single infor-mation point and makes the model able to concentrate on multiple important infor-mation points at the same time, which in turn optimizes the model performance. While these attention methods can achieve global attention to some extent, they inevitably require more parameters.

Meanwhile, Convolutional Neural Networks (CNNs) are also prevalent in the re-search of speech emotion recognition. For example, Zhao et al. in \cite{c10} combined an LSTM network and a full convolutional network (FCN), the latter extracting features by concatenating multiple 2D convolutional layers. Aftab et al. in \cite{c11} designed a lightweight FCN to extract features in-depth by increasing the number of convolu-tional layers, and Mekruksavanich et al. in \cite{c12} explored the possibility of applying a 1D convolutional method applied to Thai sentiment recognition. Although these mul-tilayer convolutional network strategies are effective in feature extraction, they usually focus only on the output of the last layer of the CNN, which may ignore the fine-grained information in the shallow layers, which limits the model's ability to capture long-term contextual information.

To address the shortcomings in the existing methods mentioned above, our study proposes a novel approach to speech emotion recognition based on the progressive channel querying technique. By combining the dual perspectives of speech and spec-trogram and introducing the channel query module, this method can model speech emotion signals dynamically in the channel dimension in a progressive manner. Our main contributions are as follows:

 \vspace{-0.4\baselineskip} % 移除前一个段落的间距
\setlength{\parskip}{0.5pt}

\setlength{\itemsep}{0pt}
\setlength{\parsep}{0pt}
\setlength{\parskip}{0pt}
\begin{itemize}[label=\textbullet]
\item 	We designed a \textbf{M}ultilayer \textbf{L}ightweight CNN (MLCNN) branch to extract feature outputs from different layers. Further, we introduce the overall framework of the PCQ network. This framework merges the MLCNN branch, using spectrogram as input, with the WavLM pre-trained model branch, using original speech as input.
\item We designed a \textbf{C}hannel \textbf{S}emantic \textbf{Q}uery (CSQ) module for querying and integrat-ing semantically similar sentiment features from neighbouring layers of an MLCNN in the channel dimension. This module exploits the temporal properties of speech signals to dynamically model speech emotion, thereby capturing the tem-poral variation of emotional information in the channel dimension. We achieve step-by-step acquisition of emotional information by integrating multiple CSQ modules in the PCQ framework.

\item Our method achieves good results on the IEMOCAP dataset and the EMODB da-taset. 

\end{itemize}

\section{Proposed Method}
This section provides a comprehensive overview of our proposed approach, as shown in Fig. 1. Section 2.1 describes the structure of the sub-network MLCNN. Section 2.2 describes the WavLM pre-trained model and the processing of its output features. Section 2.3 introduces the CSQ module. Section 2.4 outlines the overall network structure of PCQ. 

\subsection{MLCNN Branch}
As shown in Figure 1(B), the MLCNN we developed consists of four layers with \{16, 32, 48, 64\} channels per layer. For the layer setting of the MLCNN, we have explained in detail as well as ablation experiments in Section 5. As the network depth increases, the MLCNN will output deeper semantic features layer by layer. Specifically, the network contains four layers whose outputs are  $x_{1},x_{2},x_{3},x_{4} $, and the feature of the mth pair of adjacent layers is denoted as $f_{m} $. Thus, we define:$f_{1}=[x_{1},x_{2} ]$, $f_{2}=[x_{2},x_{3} ]$, $f_{3}=[x_{3},x_{4} ]$, where the $[\cdot ,\cdot ]$ symbols are used to denote the selected two neighboring layer features. In addition, each layer contains a \textbf{P}arameter-efficient \textbf{D}epth \textbf{C}onvolution (PDC) block, which consists of two point convolutions, a depth convolution, and a channel attention block, as shown in Figure 1(C). The input feature is $\chi \in \textbf{R}^{C\times W\times H }$, where C denotes the number of channels, W is the width, H is the height, and the cth channel of the input feature map is denoted as $\chi _{c} $. For each feature map, the global average pooled feature $Gap(\chi)$ is obtained by computing the average of all its elements.
$Gap(\chi )$ is then fed into a fully connected layer $\Upsilon $ and processed by a sigmoid function $\delta $ to learn the importance of the sentiment information in each channel and generate the channel weights $\omega$ accordingly. Then, by multiplying the weights  $\omega$ with the input feature $\chi $, we obtain the output feature y. In Eq.(1), $(\chi _{c})_{i,j}$ is the element with position $(i,j) $ on the cth channel.
\begin{align}
    \mathit{Gap} (\chi )& = \frac{1}{W\times H } \sum_{i=1}^{W} \sum_{j=1}^{H} (\chi _{c})_{i,j} ,  \\[8pt]
    \omega &= \delta(\Upsilon(\text{\textit{Gap}}(\chi))), \\[8pt]
    y &= \chi \otimes \omega.
\end{align}

 \begin{figure}
\includegraphics[width=\textwidth]{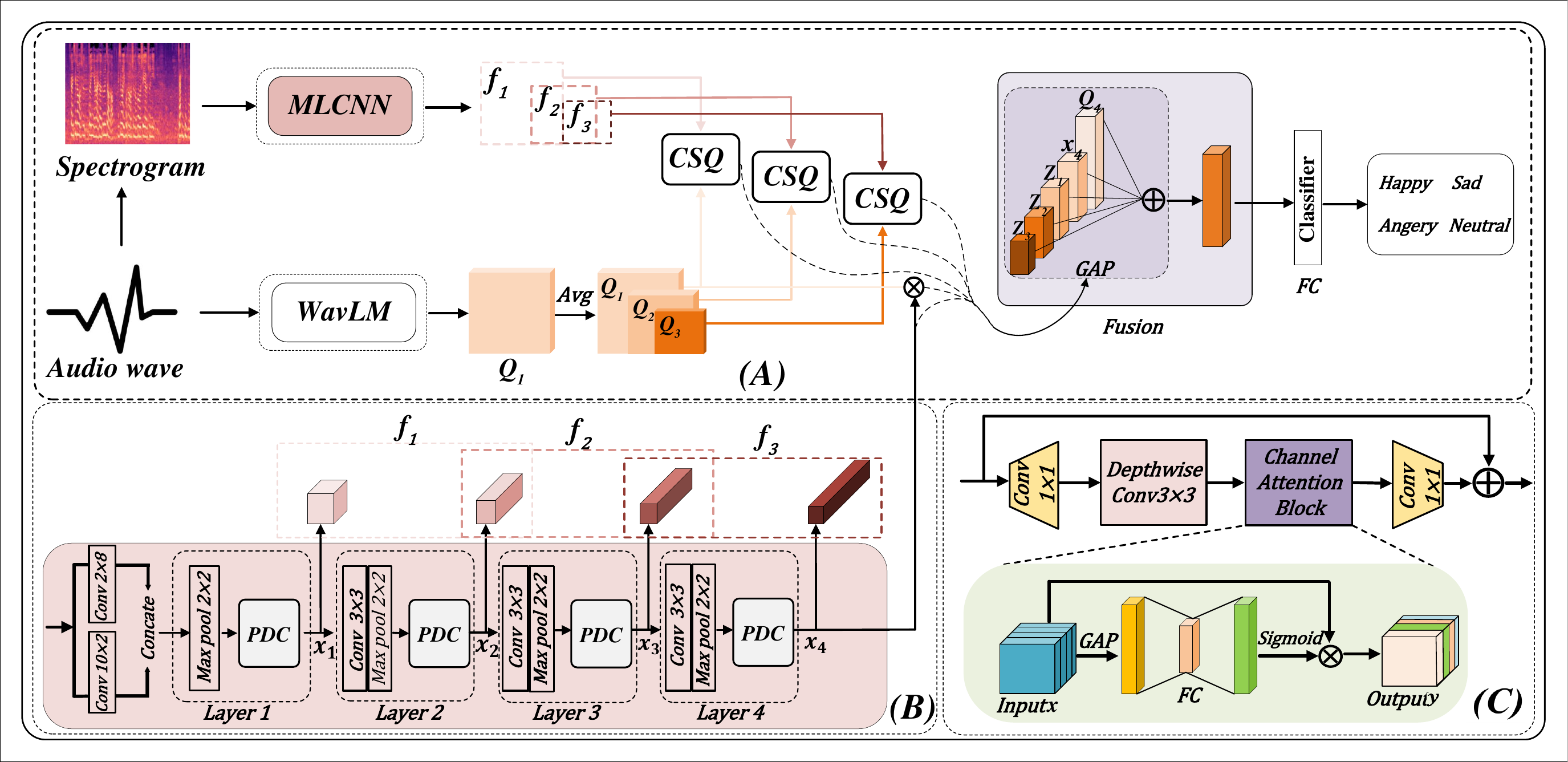}
\caption{(A) The overall PCQ framework. (B) The MLCNN network. (C) The PDC module.
} \label{换题目主图}
\end{figure}
\vspace{-0.5cm} % 减小图形和下一节标题之间的距离，根据需要调整这里的数值
The attentional mechanism therefore enables the model to understand and emphasize key emotional information in each channel more accurately. The PDC module adopts an architecture that sequentially connects dot convolution, deep convolution, channel attention mechanism, and dot convolution. This design not only significantly improves the performance of the model in capturing emotionally relevant information, but also effectively reduces noise interference.

The PDC module maintains an equal number of input and output channels, denoted as \(C\). Under equal conditions, traditional 3x3 convolutions require \(9C^2\) parameters, while the parameter count of the PDC module is \((16/3)C^2 + 18C\). From the analysis, it is found that the number of parameters of the PDC module will be lower than the traditional 3x3 convolution when \(C > 4.90\). The relevant parameter count comparisons are detailed in Table 3 in Section 4. In addition, in Table 4 in Section 4, we also show that the PDC module improves in terms of accuracy compared to the traditional 3x3 convolution.
\vspace{1mm} % 这里调整为您希望减少的量
\subsection{WavLM Branch}
As shown in (A) of Fig. 1, the WavLM pre-trained model, serving as an encoder within a branching network that takes speech signals as inputs, effectively models long-term contextual sequences. Second, in the work \cite{c13}, Zhao et al. revealed the layer-to-layer variability in the pre-trained model: the part closer to the output layer tends to contain rich task-specific information, whereas the underlying layers closer to the input capture more generic features. These bottom layers capture a wide range of features, while the top layer focuses on high-level abstract features that are closely related to a specific task. In this study, we utilized the WavLM pre-trained model for the speech emotion recognition task. Compared to the bottom layer of the model, the top layer of WavLM is more focused on modeling speech emotion information. Therefore, we selected features from the final output layer of the WavLM model and enhanced the network's overall recognition of emotions through multiple average pooling operations.
\begin{figure}[t]
\begin{center}
\includegraphics[width=0.75\textwidth]{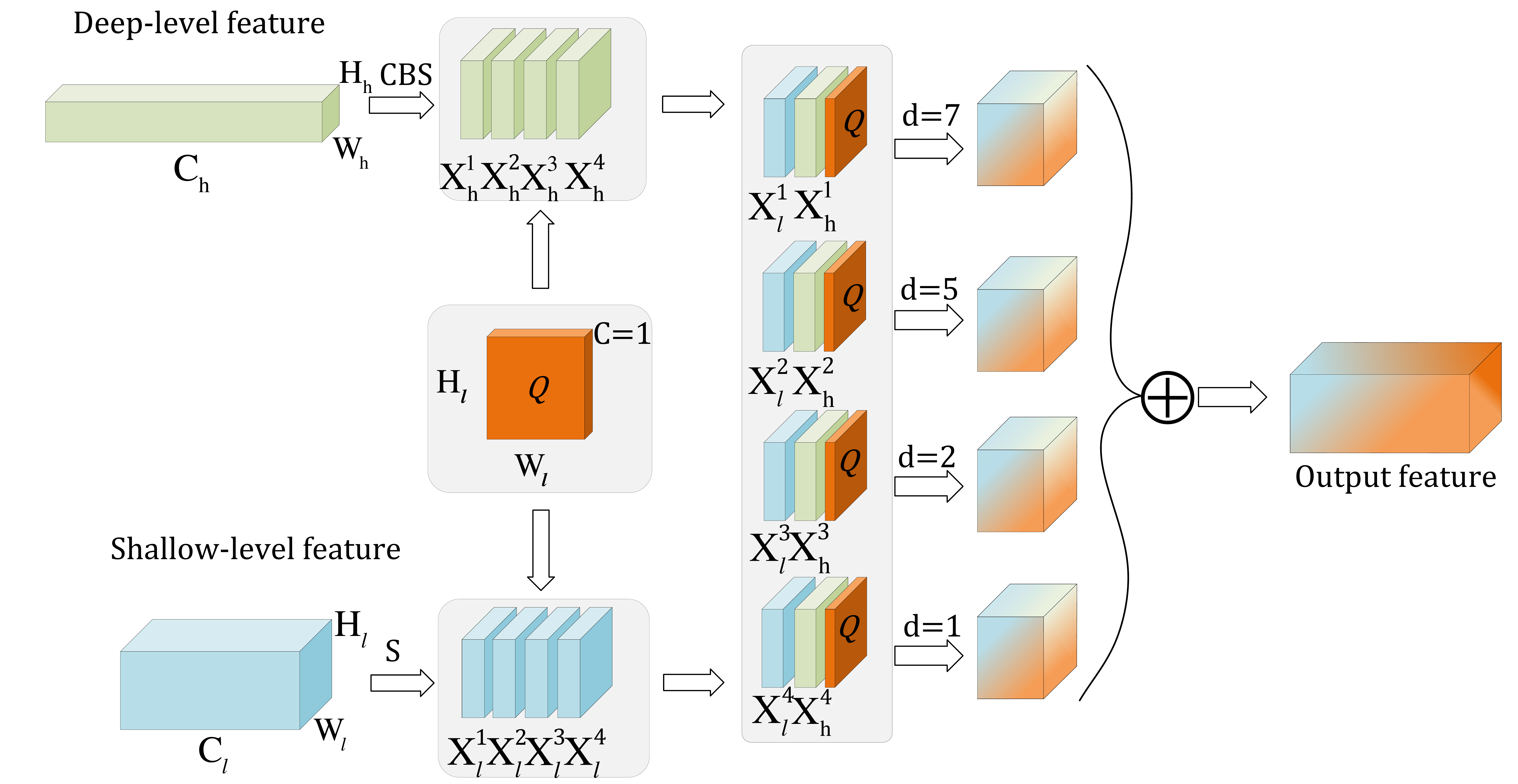}
\caption{CSQ module. The letters on the arrows in the diagram represent the following: \textbf{C} denotes that the convolution operation is first performed to match the number of channels $C_{h} $  of the high-level features with the number of channels $C_{l} $ of the low-level features; \textbf{B} denotes that the height H and width W of the feature map are adjusted to keep the high-level features ($H_{h} $,$W_{h} $) and the low-level features ($H_{l} $,$W_{l} $) the same using bilinear interpolation; and \textbf{S} denotes that the channel dimensions are segmented; \textbf{d} is the dilation rate of the convolution. } \label{CSQ}
\end{center}
\end{figure}
\vspace{-1mm} % 这里调整为您希望减少的量

\subsection{CSQ Module}
In Figure 2, we design the \textbf{C}hannel \textbf{S}emantic \textbf{Q}uery (CSQ) Module. CSQ module takes three inputs, with the first two being shallow-level features $X_l \in \mathbb{R}^{C_l \times H_l \times W_l}$ and deep-level features $X_{h}\in \mathbb{R}^{C_{h}\times H_{h} \times  W_{h} }$. To aggregate the semantically similar information on these two different feature scales, we first employ convolution and bilinear interpolation to resize the deep-level speech features to match the size of the shallow-level speech features, $C_h=C_l$, $H_h=H_l, W_h=W_l$.
Next, we evenly divide the channels of the two features into four groups (\textit{Group}= 4) for positional encoding. We introduced a channel query token $Q$ which queries and aggregates channel features with identical positional encodings, creating $\varpi^{i}\in \mathbb{R}^{(Group/2+ 1)\times H_{l} \times W_{l} }$, $i=[1,2,3,4]$. As shown in Eq. (4), in each $\varpi^{i}$, $Q$ efficiently synthesizes similar emotional information of the same positional encoding, utilizing its pre-trained knowledge. In high-dimensional channels, deeper semantic information exists. To further extract the deeper semantic feature $\eta ^{i}\in \mathbb{R}^{(Group/2+ 1)\times H_{l} \times W_{l} }$, for four blocks from different channel dimensions, we use 3x3 dilation convolution with dilation rates of $d_{i}$ = [7,5,2,1]. 

Based on the correlation between emotion and time, we concatenate \(\eta ^{i}\) to obtain $\hat{y} \in \mathbb{R}^{C_{l} \times H_{l} \times W_{l}  } $. Next, a convolution operation highlights the emotional information in  $\hat{\eta }$. 
\vspace{-2mm} % 这里调整为您希望减少的量
\begin{equation}
\setlength{\abovedisplayskip}{1pt}
\setlength{\belowdisplayskip}{1pt}
   \varpi^{i}= Concat(X_{l}^{i} ,X_{h}^{i},Q ).
\end{equation}

\begin{equation}
\setlength{\abovedisplayskip}{1pt}
\setlength{\belowdisplayskip}{1pt}
   \eta ^{i}= Conv_{3\times 3}^{d_{i} } (\varpi^{i} ).
\end{equation}
\subsection{The Proposed Overall Framework (PCQ)}
In Figure 1(A), PCQ is composed of three main components: the MLCNN sub-branch, the WavLM pre-training branch, and the CSQ module. Firstly, spectrogram features are input into the MLCNN, as illustrated in Figure 1(B). The MLCNN produces three sets of outputs, namely, $f_{1}$, $f_{2}$, and $f_{3}$. These outputs are then sequentially used as inputs for the three independent CSQ modules.
After the speech signal passes through the WavLM encoder, it first obtains the channel query token $Q_{1}$ with channel number 1. Then, $Q_{1}$ undergoes two adaptive pooling operations to obtain the other two channel query tokens, $Q_{2}$ and $Q_{3}$.
These tokens are pre-trained features with global sentiment information. Next, we need to use this pre-training information to assist the PCQ network in achieving global sentiment perception. Therefore, \( Q_1, Q_2, \) and \( Q_3 \) are input as query tokens to the three CSQ modules sequentially. Next, the CSQ modules will sequentially generate three progressive features \( z_{1}, z_{2}, \) and \( z_{3} \), which are different degrees of perception of global emotion information. $Q_{1}$ obtains its self-attention features through an attention mechanism weighted by $x_{4}$.
\begin{align}
     z_{j} &=\textit{CSQ}(f_{j} , Q_{j}),  j =1,2,3. \\[5pt]
     &\hspace{1cm} Q_{4} =x_{4} \otimes Q_{1}.
\end{align}

To enhance feature fusion, the Gap (Global Average Pooling) operation is applied to $Q_{4}$, $z_{1},z_{2},z_{3}$ and $x_{4}$. Subsequently, these fused features are fed into a classifier comprising multiple linear layers for emotion prediction. as shown in Eq. (8).These fused features are fed into a classifier comprising multiple linear layers for emotion prediction,  as shown in Eq. (9).
\begin{equation}
y_{\text{fusion}} = \text{Concat}\left(\text{Gap}(z_1), \text{Gap}(z_2), \text{Gap}(z_3), \text{Gap}(x_4), \text{Gap}(Q_1)\right)
\end{equation}
\begin{equation}
\hat{y} = \text{Classifier}\left(y_{\text{fusion}}\right)
\end{equation}
\vspace{-12mm} % 这里调整为您希望减少的量
\begin{figure}
\centering
\includegraphics[width=9cm]{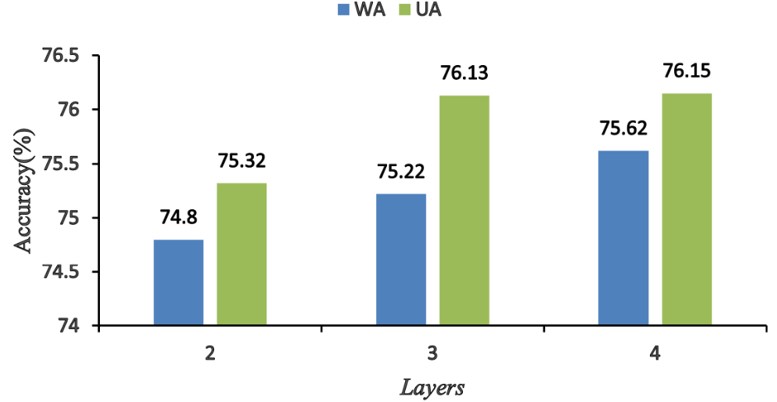}
\vspace{-5mm} % 这里调整为您希望减少的量
\caption{Visualisation of MLCNN results for different layers on the IEMOCAP dataset: blue for weighted accuracy (WA), green for unweighted accuracy (UA).} 
\label{准确度图}
\end{figure}
\begin{table}[t]
\centering
\tabcolsep=0.78cm
\caption{Comparison of baseline and state-of-the-art methods on the IEMOCAP dataset. All $(\uparrow)$ means in the experimental table indicate that higher is better, and $(\downarrow)$ means indicate that lower is better.  ($\dagger $)indicates data derived directly from the paper.}
\begin{tabular}{l|cc}
\hline
\textbf{Models}               & \textbf{WA($\uparrow$)}  & \textbf{UA($\uparrow$)}  \\ \hline
HNSD\cite{c14}$^{\dagger}$                & 70.50           & 72.50           \\
GLAM\cite{c15}$^{\dagger}$                   & 73.70          & 73.90          \\
SMW CAT\cite{c16}$^{\dagger}$                & 73.80          & 74.25          \\
GLNN\cite{c17}$^{\dagger}$                & 71.83          & 65.39          \\
AMSNet\cite{c18}$^{\dagger}$                & 69.22          & 70.51          \\
Cross-fusion\cite{c19}$^{\dagger}$          & 72.04          & 73.26          \\
MFCC+Spectrogram+W2E(base)\cite{c20}$^{\dagger}$ & 71.64          & 72.70          \\ \hline
\textbf{PCQ(ours)}            & \textbf{75.62} & \textbf{76.15} \\ \hline
\end{tabular}
\end{table}

\section{Experiment}
\subsection{Datasets}

\textbf{IEMOCAP }: This dataset is an English corpus. In total, it contains about 12 hours of audiovisual data, of which audio data has been widely used in automatic emotion recognition re-search. We identify four main emotions: anger, sadness, happiness and neutrality. Considering the unbalanced distribution of emotion categories in the dataset, we merge "happy" and "excited" into "happy". \textbf{EMODB }: This dataset is a German language speech library recorded by 10 participants (5 males and 5 females) covering seven different emotional expressions: anger, disgust, fear, happiness, sadness, surprise, and neutrality. The entire database contains about 535 audio clips, each ranging from 1 to 10 seconds long, providing a rich sound sample for emotion recognition studies.

\subsection{Experimental Setup}
In this study, we sampled the raw audio signal at 16 kHz and segmented it into 3-second segments, with underfilled segments using zero padding. The final prediction is based on the judgement of all the segments. Through a series of 40 ms Hamming windows, we generate spectrogram features. Each window was discrete Fourier trans-formed (DFT) as a frame to obtain an 800-point frequency domain signal, and the first 200 were taken as input features. In this way, we obtained a spectrogram of 300x200 size corresponding to each audio clip. To evaluate the model performance, we use both Weighted Accuracy (WA) and Unweighted Accuracy (UA) metrics and use 10-fold cross-validation to ensure that the results are reliable.Our emotion classi-fication task uses a cross-entropy loss function. Our system is implemented in PyTorch. The batch sizes are 16 and 32 for the IEMOCAP dataset and EMODB da-taset, respectively. The early stop is 20 epochs. We use the AdamW optimiser, and the learning rate is 1e-5. All experiments were conducted on an NVIDIA 4090 24G GPU.
\begin{table}[]
\tabcolsep=0.78cm
\caption{Comparison of baseline and state-of-the-art methods on the EMO-DB dataset. (${\star }$) indicates that all results are reproduced using publicly available source code and the original hyperparameters. ($\dagger $)indicates data derived directly from the paper.}
\begin{tabular}{l|c|c}
\hline
\textbf{Models}            & \textbf{WA($\uparrow$)}  & \textbf{UA($\uparrow$)}  \\ \hline
TSP+INCA\cite{c21} $^{\dagger }$                  & 90.09          & 89.47          \\

GM-TCN\cite{c22}$^{\dagger }$                     & 91.39          & 90.48          \\
LightSER\cite{c11} $^{\dagger}$                  & 94.21          & 94.16          \\
TIM-Net\cite{c23}$^{\dagger}$                    & 95.70          & 95.17          \\
MFCC+Spectrogram+W2E(base)\cite{c20}$^{\star }$  & 90.17          & 89.65          \\ \hline
\textbf{PCQ(ours)}         & \textbf{95.84} & \textbf{95.48} \\ \hline
\end{tabular}
\end{table}
\vspace{-5mm} % 这里调整为您希望减少的量
\begin{table}[]
\centering
\tabcolsep=0.60cm
\caption{Comparison of our proposed method with the baseline network in terms of parameter size on the IEMOCAP dataset.}\label{tab2}
\scalebox{0.9}{
\begin{tabular}{l|cc}
\hline
\textbf{Models }                                                                        & \textbf{Parameters($\downarrow$)}                                                    \\ \hline
MFCC+Spectrogram+W2E(base)\cite{c20}                                                    & 174.62M                                                     \\
\textbf{PCQ(ours)}                                                                      & \textbf{97.99M}                                             \\ 
\hdashline % 添加虚线
AlexNet(base)\cite{c20}                                                                 & 2.47M                                                       \\
\textbf{MLCNN(ours)}                                                                    & \textbf{0.092M}                                             \\ 
\hline
\end{tabular}
}
\end{table}

\section{Experimental Results and Analysis}
Fig. 3 shows the performance of MLCNN with the number of layers 2, 3 and 4 on the IEMOCAP dataset. The results show that both performance metrics improve signifi-cantly with the increasing number of layers, especially when the model reaches 4 lay-ers, both WA and UA metrics reach the highest, 75.62\% and 76.15\%, respectively. This trend demonstrates that increasing network depth effectively enhances model accuracy. However, to ensure comparability with the baseline AlexNet network, we decided against further increasing the number of layers, opting instead for the same network depth to facilitate effective model comparison under similar conditions. Therefore, a 4-layer MLCNN network was used for all subsequent experiments.

\subsection{Results and Comparisons}
Table 1 shows the performance of our proposed PCQ network compared to the three-branch network used in the baseline study described in [22] on the IEMOCAP dataset. Our method achieves significant improvements of 3.98\% and 3.45\% in terms of WA and UA metrics, respectively. In addition, Table 1 also shows some current state-of-the-art speech emotion recognition models. Compared to these models, our PCQ network exhibits a superior performance. On the EMODB dataset, as shown in Table 2, the PCQ network achieved high accuracy. Compared with the baseline network[20], PCQ improved 5.67\% and 5.83\% on WA and UA, respectively, further demonstrat-ing the effectiveness and applicability of the methodology. In the upper part of Table 3, we compare in detail
the total number of parameters of the benchmark network used on the IEMOCAP dataset with our newly designed PCQ network. From the data, the number of parameters of the PCQ network is reduced by 43.98\% compared to 
 \begin{table}[H]
\centering
\tabcolsep=0.6cm
\caption{ Results of the ablation study of the components of the proposed frame on IEMOCAP dataset. “w/o” denote without, “w/” denote with.}\label{tab2}
\scalebox{0.9}{
\begin{tabular}{l|cc}
\hline
\textbf{Models}                                                                                          & \textbf{WA($\uparrow$)}                                                                            & \textbf{UA($\uparrow$)}                                                                             \\ \hline 
\begin{tabular}[c]{@{}l@{}}\textbf{PCQ}\\  w/o PDC(w/ conv3×3) \\ w/o CSQ\\  \hspace{0.5cm}w/o WavLM \\ w/o WavLM(w/ W2E)\end{tabular} & \begin{tabular}[c]{@{}c@{}}\textbf{75.62}\\ 75.20\\ 74.65\\ 57.66\\ 70.19\end{tabular} & \begin{tabular}[c]{@{}c@{}}\textbf{76.15}\\ 75.81\\ 74.73\\ 57.68\\ 71.77\end{tabular} \\ \hline 
Spectrogram+W2E(base)\cite{c19}                                                                           & 70.05                                                                         & 71.30                                                                          \\ \hline
\end{tabular}
}
\end{table}
\vspace{0.01cm}  % 调整这个值来最小化距离
\begin{figure}[H]
\centering
\begin{minipage}[b]{1.0\linewidth}
  \centering
  \includegraphics[width=10cm]{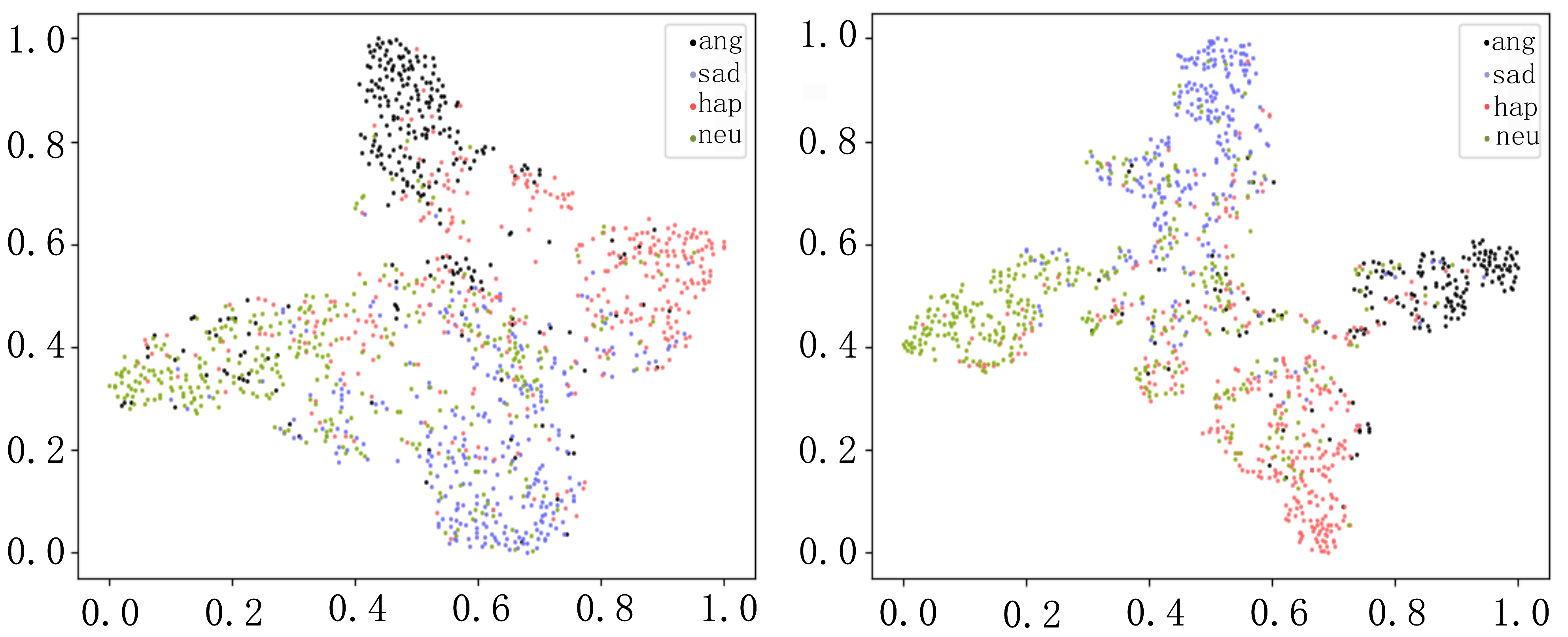}
  \centerline{(a) w/o CSQ\hspace{15em}{(b) PCQ}}
  \vspace{0.01cm}  % 调整这个值来最小化距离
\end{minipage}

\begin{minipage}[H]{1.0\linewidth}
  \centering
  \includegraphics[width=10cm]{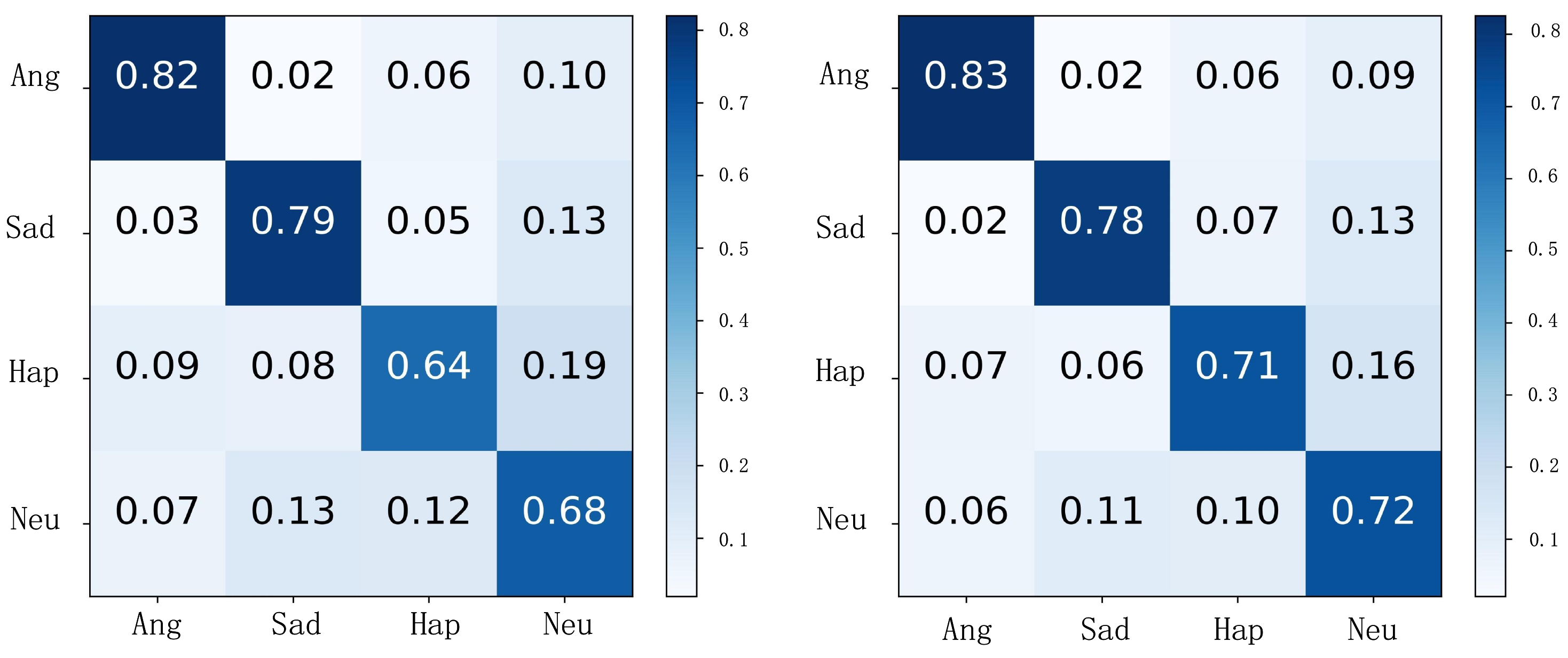}
  \centerline{(a) base\hspace{15em}{(b) PCQ}}
  \vspace{0.01cm}  % 调整这个值来最小化距离
\end{minipage}
\caption{(Top) The t-SNE visualization of feature distribution on the IEMOCAP dataset. (Bottom) Comparison of normalized confusion matrices on the IEMOCAP dataset.}
\label{fig:res}
\end{figure}
the benchmark network. In the lower part of Table 3, we compare the branching networks designed based on spectrograms. Compared to the baseline AlexNet, our designed MLCNN substantially reduces the number of parameters, which fully proves the lightweight feature of our proposed PCQ network and MLCNN network.

Table 4 details the results of the ablation experiments for each component of the PCQ network on the IEMOCAP dataset. Replacing the PDC module with the 3x3 Conv2d resulted in a reduction of WA and UA by 0.42\% and 0.34\%, respectively, despite an increase in the number of network parameters. This clearly demonstrates the advantages of the PDC module in terms of lightweight and efficiency. Without the CSQ module, the WA and UA of the network decreased by 0.97\% and 1.42\%, re-spectively. As shown in the third-to-last row of Table 4, without the CSQ module and the WavLM branch, the training accuracy of the network is significantly reduced. While the WA and UA when using Wav2Vec 2.0 as the pre-training encoder increase by 0.14\% and 0.47\% respectively compared to the baseline two-branch network, they are still 5.43\% and 4.38\% lower compared to our PCQ network. This further demon-strates that both our CSQ model and the WavLM pre-training model are indispensable key components in the PCQ backbone network. As shown in Fig. 4, the t-SNE visual-isation results on the IEMOCAP dataset show that the PCQ method with the integrated CSQ module has clearer classification boundaries than the network without the CSQ module. Furthermore, in Fig. 5, the normalised confusion matrix shows that the PCQ method significantly improves the recognition of "happy" and "neutral" emotions on the IEMOCAP dataset.
\vspace{-5mm} % 这里调整为您希望减少的量
\section{Conclusion}
In this study, we propose a new framework for speech emotion recognition named Progressive Channel Querying(PCQ). The method mainly queries and integrates similar sentiment features in the channel dimension through the CSQ (Channel Semantic Query) module. Applying the CSQ module at different layers in the PCQ framework enables a gradual enhancement of the understanding of the sentiment information, thus allowing the model to acquire sentiment features in a progressive manner. Experimental results on the IEMOCAP dataset and the EMODB dataset show that our method achieves significant improvements on the SER task compared to existing techniques. SER tasks are used in a particular scenario. For future research, we will work on multi-scene SER research, i.e., multimodal emotion recognition, where the input is speech, image, text, and other modalities.

\textbf{Acknowledgements.} The following projects jointly supported this work: the Tianshan Excellence Program Project of Xinjiang Uygur Autonomous Region, China (2022TSYCLJ0036), the Central Government Guides Local Science and Technology Development Fund Projects (ZYYD2022C19), and the National Natural Science Foundation of China (62303259), the Graduate Student Research and Innovation Program in the Xinjiang Uygur Autonomous Region (XJ2024G089).


\bibliographystyle{splncs04}
\bibliography{refer}
\end{document}